# Liquid-to-liquid phase transition underlying the structural crossover in a supercooled metallic liquid

S. Lan[a], M. Blodgett[b], K.F. Kelton[b], and X.-L.Wang[a,1]

[a]Department of Physics and Material Science, City University of Hong Kong

83 Tat Chee Ave., Kowloon, Hong Kong

[b]Department of Physics, Washington University

One Brookings Drive, St. Louis, MO 63130-4899, USA

[1]To whom correspondence may be addressed: xlwang@cityu.edu.hk

**Abstract**

The existence of a "crossover region" in glass-forming liquids has long been considered as a general phenomenon that is as important as the glass transition (1-3). One potential origin for the crossover behavior is a liquid-to-liquid phase transition (LLPT). Although a LLPT is thought to exist in all forms of liquids (4), structural evidence for this, particularly in supercooled liquids, is scarce, elusive, and in many cases controversial. A key challenge to the search for a LLPT in a supercooled liquid is the interference of crystallization during cooling. Crystallization induces major structural changes, which can overwhelm and therefore mask the more subtle changes associated with a LLPT. Here, we report the results of an in-situ containerless synchrotron study of a metallic-glass-forming liquid ($Zr_{57}Nb_5Al_{10}Cu_{15.4}Ni_{12.6}$) that show distinct changes in the liquid structure at ~1000 K, a temperature well below the melting temperature of the liquid and 150 K above the crystallization temperature. The structural transition is characterized by growing short- and extended-range order below the transition temperature, and is accompanied by a concurrent change in density. These data provide strong evidence for a LLPT in the supercooled metallic liquid, particularly in the light of a recent computer simulation study (5). That a LLPT is found in a metallic liquid supports the increasingly widely held view that a LLPT may be a common feature of a variety of liquids.

The existence of a Liquid-to-Liquid Phase Transition (LLPT) has been a subject of intense debate within the liquid and glass communities. Experimental evidence for LLPTs has been found in only a few, primarily covalently bonded, liquids. Among these, the clearest evidence comes from high-pressure experiments. For example, a sharp pressure-induced first-order LLPT was demonstrated in phosphorus by in-situ synchrotron X-ray diffraction (6). Separately, a neutron diffraction experiment revealed a crossover from a low-density to high-density form of water, also under applied pressure (7). However, other than when induced by high pressure, evidence for a LLPT in a supercooled liquid is far less clear. The onset of crystallization limits the amount of possible supercooling and the structural changes that accompany crystallization obscure the more subtle changes associated with a LLPT. A recent letter by Tanaka et al. (8) discussing triphenyl phosphate (TPP) serves to illustrate this point. Supercooling of TPP leads to a density change, but density data alone cannot reveal whether this is due to a LLPT, or to the formation of nano-crystals, a controversy that has lasted more than a decade. Using light scattering, Tanaka et al. showed that, in fact, both crystallization and LLPT occurred in their experiment, with the LLPT likely catalyzing the crystallization.

In parallel with experimental investigations, computer simulations have also been extensively used to explore structure and phase stability in supercooled liquids. Here too, the results can be marred by crystallization if the system is not cooled fast enough, leading to incorrect interpretations. This controversy was highlighted by a recent paper simulating structural evolution in deeply cooled water (9). In contrast to earlier studies, a LLPT in water was produced with the ST2 potential model and advanced sampling. The study showed that two metastable liquid phases and a stable crystal phase coexist in the same supercooled liquid, and that at room temperature the system is able to fluctuate freely between the two liquid phases, rather than crystallize, due to the slow kinetics.

Thus, in spite of the extensive studies devoted to this topic, whether or not a LLPT occurs in a supercooled liquid, and if so, the nature of the LLPT, remain unsettled issues. The containerless processing of liquids is a powerful experimental method to suppress crystallization, allowing access to the more deeply supercooled liquid region. A recent electrostatic levitation experiment by Li et al (10). suggested the possible occurrence of a LLPT in a metallic glass-forming liquid, based primarily on the observed hysteresis of the specific volume during heating and cooling. Subsequent studies by Busch et al. (11) and Kelton et al. (12) showed that the hysteresis was actually due to slow crystallization. However, Li et al also noted a change in the slope of the temperature dependence of the specific volume during cooling. This could be a hint pointing to a LLPT, which, like the case of TPP (8), may be obscured by the subsequent crystallization. Motivated by this observation, we designed a synchrotron X-ray diffraction experiment to search for the ultimate structural evidence for a possible LLPT. There were two experimental challenges: (1) the ability to levitate a high temperature liquid under high vacuum, while on the synchrotron beam line and (2) an X-ray beam line of sufficiently high intensity to allow quality data to be obtained at an approximately one Hz or faster data rate in order to capture the structure evolution during cooling. These requirements were met by using an electrostatic levitation facility constructed at Washington University in St. Louis (WU-BESL) on the 6-ID-D beamline at the Advanced Photon Source, Argonne National Laboratory (13). The chosen sample was VIT 106 ($Zr_{57}Nb_5Al_{10}Cu_{15.4}Ni_{12.6}$), one of the bulk metallic glass alloys studied by Li. et. al. (10) and an excellent glass former. Below, we describe the structural changes observed during cooling to $T_g$ that provide strong evidence for a LLPT in the supercooled liquid, particularly when viewed in the light of recent molecular dynamics (MD) simulations (5).

**Results**

Fig. 1A shows the temperature of a levitated VIT 106 sample during free cooling in the electrostatic levitator. The absence of a sudden rise in temperature (recalescence) in the temperature-time cooling curve demonstrates that the bulk of the sample was able to supercool through $T_g$ (~ 682 K) to form a glass. However, the appearance of small broad diffraction peaks below T ~ 850 K indicates the formation of a small amount of nano-crystals, which will be discussed further below.

Fig. 1B shows the specific volume measured during cooling. The scatter of the specific volume data near $T_g$ was due to overcharging (see the Methods section). A small discontinuity is indicated by the red arrow in the slope of the specific volume data; the slope becomes larger at T ~ 980 K, but reverts back to its original value at T ~ 915 K. We repeated the specific volume measurements several times. Similar discontinuities were detected which are consistent with Li's observations (10).

Fig. 1C shows the total structure factor S(Q) at selected temperatures. The diffraction data are characteristic of metallic liquids and glasses, with a first sharp diffraction peak (FSDP), followed by a second peak and a shoulder. The positions of these peaks are denoted as $Q_1$, $Q_{21}$, and $Q_{22}$, respectively. A rather dramatic change is seen in the evolution of the peak height for $Q_{21}$, which is illustrated in Fig. 1D. The peak height shows a linear increase with temperature on cooling (due to the Debye-Waller factor) until ~ 1000 K, where a distinct upturn is observed. The temperature of this upturn (T* ~ 1000 K) is also quite close to the red arrow in Fig. 1B, where a small change in slope is observed. There is no evidence of crystallization at $T^*$, even in the difference curves for successive S(q)s, which are sensitive to as low as $10^{-6}$ volume fraction transformed (see Fig. S1 in SI Appendix). Another upturn is seen at T ~ 850 K. Close examination of the diffraction data confirms that this is where a small fraction of nano-crystals are first formed (see Fig. S2 in SI Appendix). Further examination of the Time-Temperature-Transformation (TTT) diagram for this alloy confirms that T* cannot possibly be due to crystallization (14). The nose temperature of the TTT diagram is 875 K. Above this temperature, the incubation time for the crystalline phase increases steeply with temperature. At 950 K, the incubation time is already ~$10^5$ seconds. This can be compared to the length of our entire experiment, which lasted no more than 300 seconds. Similarly, phase separation in the liquid state, in which an unstable liquid decomposes into two (or several) forms of liquid with different chemical compositions, can also be ruled out. Phase separation is usually a precursor of crystallization (15, 16). For instance, simultaneous diffraction and small angle scattering experiments demonstrated (17) that in the supercooled liquid of VIT105, phase separation occurred immediately before nanoscale crystallization. Thus, the time scales for phase separation and crystallization are expected to be comparable. Given that crystallization and, for that matter, phase separation at T* would take $10^5$ seconds, the observed structural changes therefore cannot be due to phase separation.

In the following, we show, through a detailed analysis of the diffraction data, that the observed structural changes at $T^*$ are characteristic of a LLPT. We chose to start our analysis from the first sharp diffraction peak (FSDP), which characterizes the ordering beyond that of the nearest neighbors. The FSDP is asymmetric and its shape is rather difficult to define. To avoid the subjectivity of an assumed function for the peak shape, the first- and second- moments are used to characterize the FSDP's position and width, respectively; these are shown in Fig. 2A-B. As might be expected, upon cooling the position of $Q_1$ shifts linearly with temperature to higher values, corresponding to a decrease in the specific volume and an increase in the density. Again, a significant deviation from the linear relationship begins to develop at T*, and another upturn in $Q_1$ occurs at ~ 850 K, the temperature corresponding to the formation of a small fraction of nano-crystals. To further ensure that T* is not due to crystallization, the integrated intensity was calculated over a Q-range of 3.0347-3.0601 Å$^{-1}$, a region where prominent diffraction peaks from the crystalline phase are located. As can be seen in Fig. 2A, crystallization occurs only below $T_x$ ~ 850 K.

The evolution of the second moment also shows two features, first at T* and then at $T_x$. This second moment is related to the peak width which provides a measure of the correlation length. The reduction of the peak width at T* signifies an increasing correlation length below T*. In lieu of the peak width, the peak height could also be used as an approximate measure of the correlation length. The evolution of the peak height for $Q_1$ is shown in Fig. 2C. An increase in peak height is clearly seen at T*, followed by another increase at $T_x$.

The position of $Q_1$ can be scaled to the specific volume. Ma et al. have shown that the molecular volume $v_a \propto Q_1^{-D}$, where D=2.31 is the fractal dimension of metallic glasses (18). A similar relationship was later found by Gangopadhyay et al. for metallic liquids (19), with essentially the same D-values (2.28 – 2.31). We have chosen to use D=2.31 and superimposed the results of $Q_1^{-D}$ in Fig. 1B. As can be seen, $Q_1^{-D}$ decreases continuously during cooling, but the slope vs. temperatures shows a clear cross over at T*, which matches a small kink found in the temperature dependence of the specific volume.

Having analyzed the extended-range ordering embodied by $Q_1$, we now turn to an analysis of the short-range order. This is achieved by examining the reduced pair distribution function G(r), obtained from the Fourier transform of S(Q). Fig. 3A shows the G(r) at selected temperatures during cooling. The first peak in G(r) corresponds to the nearest neighbor shell, and therefore can be used as a crude measure of the short-range order. The integrated intensity in G(r) over 2.37-3.31 Å (underneath the first peak) is shown as a function of temperature in Fig. 3B. Like in the S(q) data, the integrated intensity of G(r) shows an upward deviation from the linear relationship below T*, indicating an enhanced short-range ordering in the low-temperature supercooled liquid.

The G(r) shows a growing shoulder on the $2^{nd}$ peak (at r~5.69 Å) during cooling, as marked by the arrow in Fig. 3A, indicating the growth of the extended-range order beyond the nearest neighbor shell. Of note, the height of the shoulder starts to grow only when the temperature is below T*. To illustrate this point, the integrated intensity of G(r) over the r-range of 5.69-5.70 Å is superimposed in Fig. 3B. The integrated intensity is essentially unchanged on cooling until T* is reached, but below T*, there is a sharp increase. These results indicate that fundamental clusters, featured by the first peak in G(r) (up to r~4 Å), are becoming increasing interconnected in the liquid phase. A recent MD study of $Cu_{64}Zr_{36}$ finds evidence of this behavior (5); here, it is demonstrated experimentally for the first time. The integrated intensity of G(r) over a larger r-range of 15.97-17.01 Å is shown as a function of temperature in Fig. S3 of SI Appendix, further demonstrating the growth of extended-range order below T*.

**Discussion**

The results of this study provide direct structural evidence for a LLPT in the supercooled liquid region. Microscopically, the phase transition is characterized by growing order at multiple length scales below the transition temperature T*. In metallic glasses, the fundamental building blocks are solute-centered clusters, with solute or minor atom at the center surrounded by solvent or majority atoms (18, 20-22). It is generally recognized that none of the clusters are perfect - they are plagued by chemical and topological disorder (18). Near $T_l$, chemical and topological ordering begins to develop cooperatively and eventually percolates at the glass transition temperature, $T_g$, below which the amorphous material is kinetically frozen. This is demonstrated by MD simulations for Zr-Cu (5, 23, 24) and for Zr-Cu-Al metallic glasses (23), with atomic clusters of icosahedral symmetry. Our data, summarized in Fig. 3B, showed that indeed, a LLPT is accompanied by enhanced short-range ordering. Furthermore, Fig. 3B and Fig. S3 indicates that below T*, there is also ordering on a longer scale (medium-range ordering and beyond). Based on a recent MD simulation, this would correspond to local clusters in the liquid becoming increasingly connected. Between T* and $T_x$, the integrated intensity of G(r) over the r-range of 5.69-5.70 Å increased by almost a factor of two. The sharp increase suggests that this quantity may well characterize the order parameter of the LLPT. The increased correlation length is also reflected in the temperature dependence of $Q_1$, where a distinct decrease is observed in the $2^{nd}$ moment of the peak, see Fig. 2B.

Our experimental observations support the depiction of LLPT by current theories. Consider, for example, Tanaka's theory (4, 25) for a LLPT which is predicated on two order parameters: density and bond ordering. The density order parameter maximizes the density (or packing) to lower the attractive interaction energy, which ultimately leads to long-range ordering or crystallization. The bond order parameter on the other hand, describes the packing of "locally favored structures," which captures both short- and medium-range order. In this view, the liquid is in an excited state at high temperatures. Upon cooling, the liquid transforms into an energetically more favorable state with higher density.

A similar picture was proposed by Kivelson et al. (1, 26), arguing that locally preferred structures form upon cooling, but with a symmetry that cannot fill space. The incompatibility between local order and global space filling forms the basis for the concept of frustration (of long-range order). This view is in fact consistent with fractal packing of solute-centered clusters found in a variety of metallic glasses (18, 27). As a result of frustration, the locally preferred structures fluctuate cooperatively or collectively over a limited spatial scale, which characterizes the medium-range order.

An example of a locally favored structure that could not fill space is icosahedral order, which was first proposed by Frank (28) and later confirmed experimentally in metallic liquids (29, 30) and glasses (31, 32). Frank initially proposed icosahedral order to explain the nucleation barrier in metallic liquids, since it is incompatible with translational periodicity. However, MD simulations by Pederson et al. (33) for a binary metallic liquid showed that this needs not be the case. They demonstrated that the short-range order in their supercooled liquid could adopt the same short-range order (Frank-Kasper cluster in this case) as in the crystal. The supercooled liquid is stabilized instead by the competition of multiple possible stable arrangements of crystal-like short-range order over intermediate length scales, expanding the well-known "confusion principle" proposed for glass formation (34) to include structural motifs in the liquid.

The physical origins for a LLPT have been explored in a number of simulation studies, notably in water. A LLPT in water emerges naturally with a Jagla potential (35) of two length scales that are characterized by an attractive and a repulsive ramp, respectively. A more recent study with an ST2 potential model also produced a LLPT in water (9). Two metastable liquid phases and a stable crystal phase are found to coexist in the same supercooled liquid. The liquid state is stabilized by kinetics, and the system is able to fluctuate freely between the two liquid phases before crystallization takes place.

Recent MD simulations by Egami et al. (36) for metallic liquids suggest that elementary excitations in high temperature metallic liquids are local configurational changes in the atomic connectivity network. These excitations, involving changes in the local coordination number, are the elementary steps to change the atomic connectivity network, and directly control the macroscopic viscosity. A further MD simulation of $Cu_{64}Zr_{36}$ (5) expanded this study and has identified a temperature $T_D$, located at approximately 1.4 $T_g$, where the diffusion coefficients for both Cu and Zr decrease abruptly and the lifetime of the interconnected clusters becomes longer than the Maxwell time for viscous flow. In essence, the ordered regions begin to "solidify" below this temperature. An estimate for $T_D$ for VIT 106 leads to a temperature remarkably close to the $T^*$ found in our experiment, below which the atomic clusters are shown to become increasingly connected. The simulation by Pederson et al. (33) with Frank-Kasper clusters in metallic liquids also showed that growing fluctuations of the Frank-Kasper clusters can lead to a liquid-to-liquid like structural crossover.

In-situ levitation experiments with synchrotron X-ray have been carried out before, e.g., by Wei et al. (11) on multicomponent $Zr_{41.2}Ti_{13.8}Cu_{12.5}Ni_{10}Be_{22.5}$ and Georgarakis et al. (37) on ternary $Zr_{60}Cu_{40}Al_{10}$. Evidence of structure ordering was noted, based primarily on the shift in $Q_1$ peak position. In both cases, however, the peak widths of $Q_1$ are shown to exhibit an increase, rather than a decrease, below an assumed transition temperature, These observations are in conflict with current theories (1, 2, 4, 25, 26, 38-40) and computer simulations for LLPT (5, 33, 36), which predict increasingly connected clusters and stronger fluctuations at lower temperatures. In contrast, all aspects of our measurements are consistent with current views of liquid ordering and predictions by MD simulations.

It may be tempting to associate the observed LLPT with the crossover predicted by the Mode-Coupling Theory (MCT) (10, 39). In MCT, the crossover temperature $T_c$ is determined by scaling the diffusion data as a function of temperature, $D \sim [(T-T_c)/T_c]^\gamma$ (40). For Zr-based metallic liquids $T_c \sim 870$ K (41), which is more than 100 K lower than T* observed in the present experiment and in Li's work (10). Given this large difference, it is unlikely that the observed LLPT is related to the MCT crossover.

**Conclusions**

In summary, fast in-situ synchrotron measurements coupled with electrostatic levitation has enabled the direct observation of a structural crossover in a supercooled metallic liquid. The dramatic changes in short- and medium-range order provided strong evidence that the observed structural crossover is linked to a LLPT, an experimental observation that is in remarkable agreement with recent MD simulations (5). Taken together, these results demonstrate that below the transition temperature, the atomic clusters develop strong chemical and topological ordering, and the clusters also become increasingly connected. The findings of our study will make an important contribution to fundamental understanding of the universal feature of LLPT in supercooled liquids and, in practical sense, the development of metallic glasses with superior glass forming abilities.

**Materials and methods**

Specimen Preparation: : ~ 1g master-ingots of VIT 106 alloy (composition $Zr_{57}Nb_5Al_{10}Cu_{15.4}Ni_{12.6}$ with high-purity raw materials) were first prepared by arc-melting on a water-cooled Cu hearth in a Ti-Zr-gettered high-purity argon atmosphere (99.998%). In order to make sure that the composition was homogeneous, each ingot was melted at least three times. During alloying, the arc-melting chamber was first evacuated to less than 50 mtorr and then refilled with Ti-Zr-gettered high-purity argon. This process was repeated several times for each melting to ensure a low oxygen concentration in the chamber. The mass loss of the ingot after arc-melting was strictly controlled to less than 0.05%. The master ingot was broken into small portions and approximately spherical ESL samples of mass 50-60 mg were produced by arc-melting these small portions for the synchrotron experiment.

Levitation and heating: The Washington University Beamline ESL (WU-BESL) was incorporated into the high-energy synchrotron beamline 6-ID-C of the Advanced Photon Source at Argonne National Laboratory. The spherical samples were levitated inside the WU-BESL chamber under high vacuum $\sim 10^{-7}$ Torr and heated to ~300 K above the liquidus temperature using a 50 W diode laser. The levitated sample was quickly heated to 1678 K (well above the $T_l$, ~ 1115 K) and held for a period of time to homogenize the temperature before the laser power was turned off for free cooling. The sample became overly charged from exposure to the x-ray beam between ~ 723 K and ~ 693 K, leading to significant instability (and noise in the temperature measurement). To compensate for the charge gain, a high-intensity UV light source was aimed at the electrodes of the ESL to direct electrons back onto the sample. The violent oscillations caused the sample to slightly deform as it was cooled. These deformations were frozen at the glass transition, causing the significant increase in scatter of the measured density below $T_g$.

Synchrotron scattering experiments: High-energy x-rays (129 keV, 0.0958(6) Å) was used in a transmission geometry to measure the structural evolution of the levitated spherical VIT 106 droplet during free-radiative cooling. A GE revolution 41-RT area detector was used for recording the scattering intensity with a range of $0.8 \leq Q \leq 14$ Å$^{-1}$, where Q is the magnitude of momentum transfer (details can be found in Ref. (42)). The experimental data were recorded every second. A levitated silicon sample was used to determine the sample-to-detector distance and its orientation relative to the beam normal. By using in-house software_ENREF_38(43) to correct the Compton and multiple scattering, and absorption for a spherical sample geometry, the total structure factor, S(Q), was obtained as a function of temperature. Then the pair distribution function, G(r), was calculated by taking a Fourier transform of S(Q). A two-color pyrometer at high temperatures (900-2600 K) and a single-color pyrometer at low temperature (450-1050 K) were used for temperature measurements. The specific volumes of the samples were also measured as a function of temperature from video images(44).

Calculation of the first and second moments: The first moment represents the peak position, which was calculated by the formula $Q_1 = \sum S(Q) \cdot Q / \sum S(Q)$. The 2$^{nd}$ moment describes the variance of the peak distribution, which was calculated by the formula $\sigma^2 = \sum S(Q) \cdot (Q-Q_1)^2 / \sum S(Q)$, where $Q_1$ is the first moment.

**Acknowledgements**

XLW thanks W. L. Johnson for discussions alluding to his work in ref. 10. The research at Washington University was partially supported by the National Aeronautics and Space Administration (NNX10AU19G) and the National Science Foundation (DMR 12-06707). The use of the Advanced Photon Source, an Office of Science User Facility operated by the U.S. Department of Energy (DOE) by Argonne National Laboratory, was supported under DOE contract DE-AC02-06CH11357. The assistance of D. Robinson with the X-ray measurements is gratefully acknowledged.

**Figure captions**

**Fig. 1.** (A) Cooling history of a levitated VIT 106 sample which formed a glass in the levitator. The liquidus temperature $T_l \sim 1115$ K and the glass transition

temperature $T_g \sim 682$ K, measured by DSC at a heating rate 0.33 K/s, are indicated by arrows. The temperature data with larger noise between ~723 K and ~ 693 K is due to recharging of the sample. (B) Specific volume as a function of temperature measured during cooling. The blue triangles are raw data, whereas the red dots are the smoothed data. The dark green circles are the volume calculated by scaling $Q_1$, the position of the first sharp diffraction peak in S(Q). Data noise increased below ~ 723 K because the sample was unstable and deformed slightly, as discussed in the text. A small kink can be seen as indicated by a red arrow at ~ 980 K, which is consistent with the transition temperature, T*, exhibited by $Q_1^{-2.31}$ data during cooling. As shown by the red dots, T* is at ~ 1000 K. (C) Total structure factor S(Q) at selected temperatures. The X-ray scattering data were acquired during cooling from 1389 K to 577 K with a sampling rate of 1 Hz. The X-ray shutter had to be closed temporarily for recharging between ~ 723 K and ~ 693 K, resulting in a loss of data in this temperature region. The arrows mark the first three diffraction peaks ($Q_1$, $Q_{21}$, $Q_{22}$) in the S(Q) curves. The red dotted line near $Q \sim 3$ Å$^{-1}$ identifies a small crystalline peak which developed below ~ 850 K. (D) The temperature dependence of $Q_{21}$ peak height during cooling. Two transition temperatures are readily identified, at T*~1000 K and $T_x \sim 850$ K, respectively.

**Fig. 2.** Evolution of the first moment (A), the second moment (B), and the height (C) of the $Q_1$, first sharp diffraction peak in S(Q). The dark green circles in (A) show the integrated intensity of the crystalline peak identified in FIG. 1(C) as a function of temperature. The integration Q range is 3.0347-3.0601 Å$^{-1}$. The red and green lines are linear fits of data above T*.

**Fig. 3.** (A) The reduced pair distribution function G(r) of the metallic liquid during cooling. The red arrows indicate the first, second, and seventh coordination shells. The black arrow identifies a shoulder peak that formed below T*. (b) The integrated intensities of the shoulder peak (green circles) and the first peak (blue triangles) as a function of temperature during cooling. The integration range is 5.69-5.70 Å for the shoulder peak and 2.37-3.31 Å for the first peak.

**Fig. 1.**

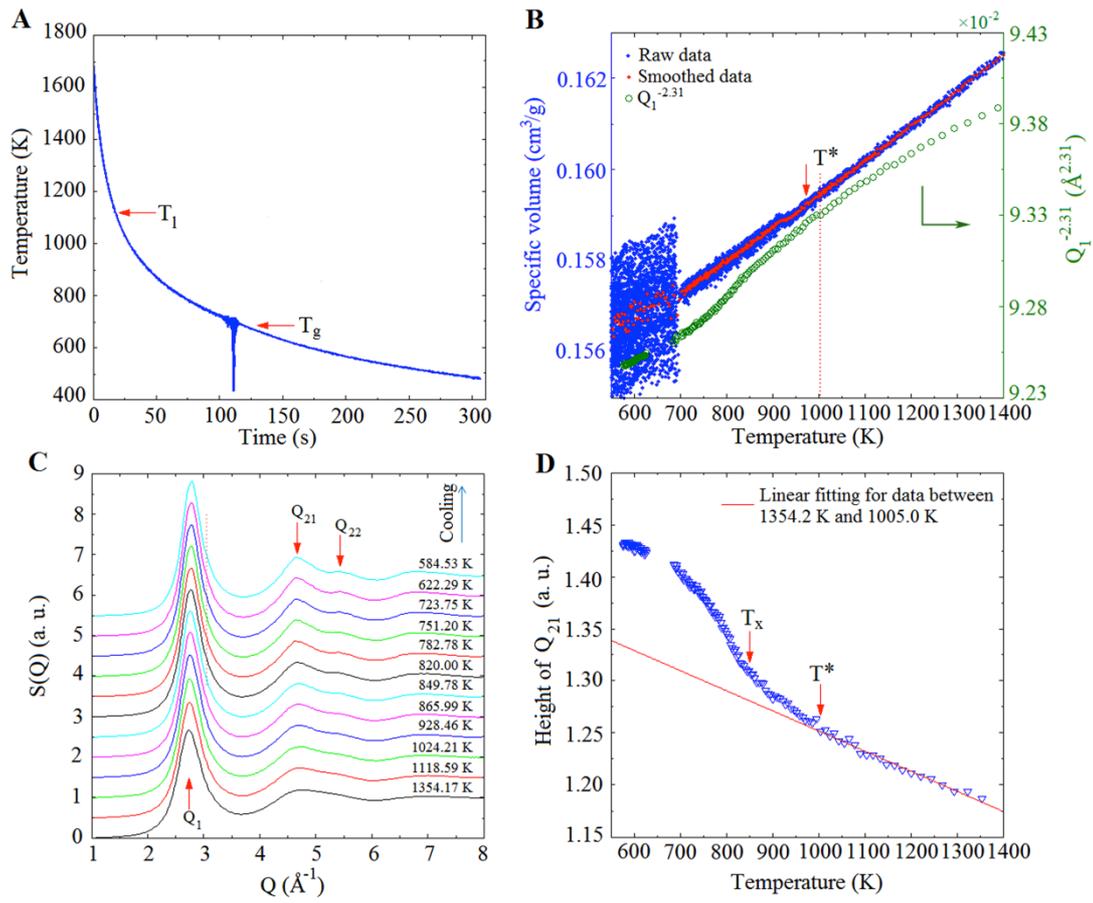

**Fig. 2.**

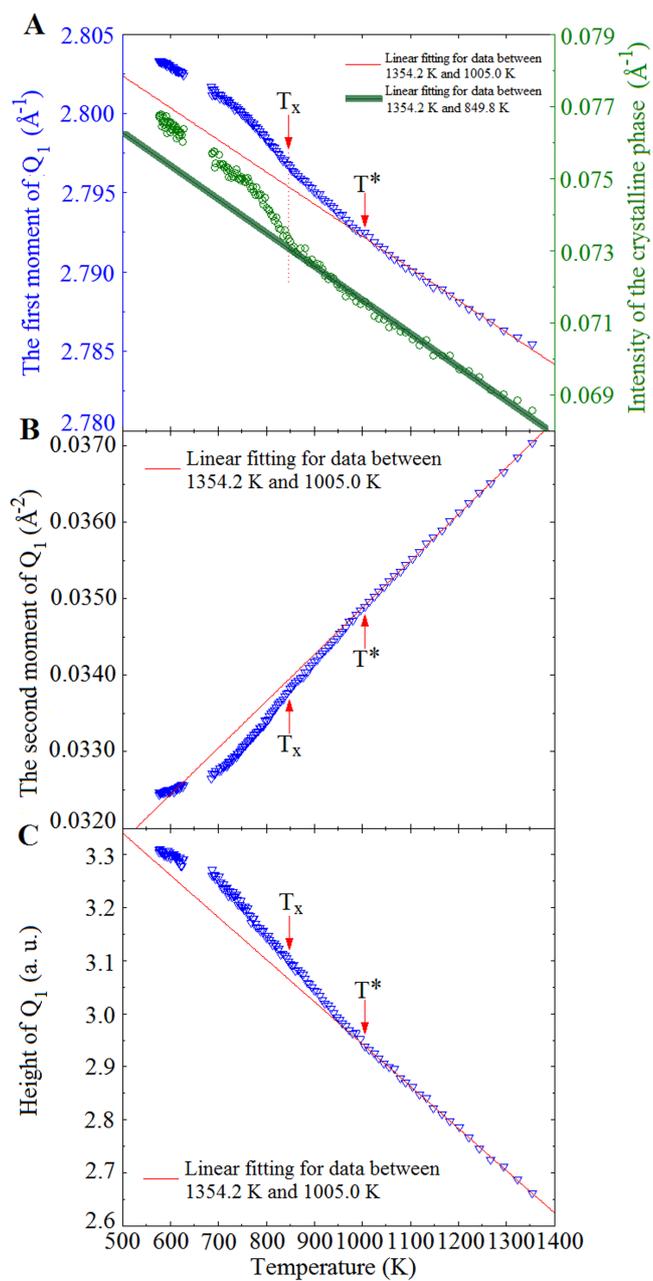

**Fig. 3.**

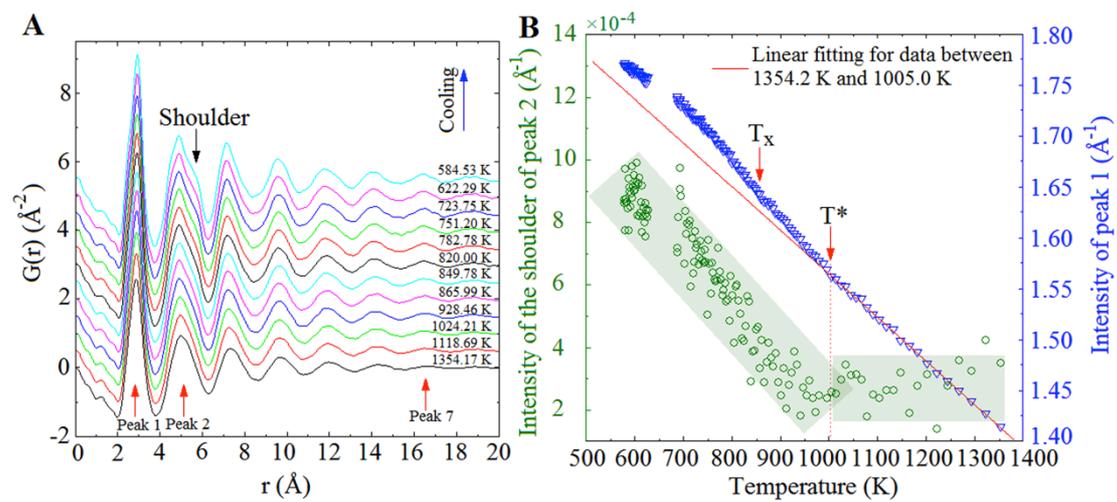

**Supporting Information**

by Lan et al..

To ascertain that the observed structure changes at T* ~ 1000 K is not from crystallization, we examined the difference curves for successive S(q)s, which are sensitive to as low as $10^{-6}$ volume fraction transformed. The difference curves, $\Delta S(Q)$, were obtained by subtracting the S(Q) at 1024 K. As shown in Fig. S1, we can clearly see that a shoulder, indicated by the red arrow, starts to develop below ~ 850 K. Fig. S2 shows the integrated intensity of the identified crystalline peak as a function of temperature. The integration was over a Q-range of 3.0347-3.0601 $\text{Å}^{-1}$. Fig. S2 clearly demonstrates that the crystallization started to take place below ~ 850 K. The integrated intensity of G(r) over a larger r-range of 15.97-17.01 Å is shown as a function of temperature in Fig. S3, further demonstrating the growth of extended-range order below T*.

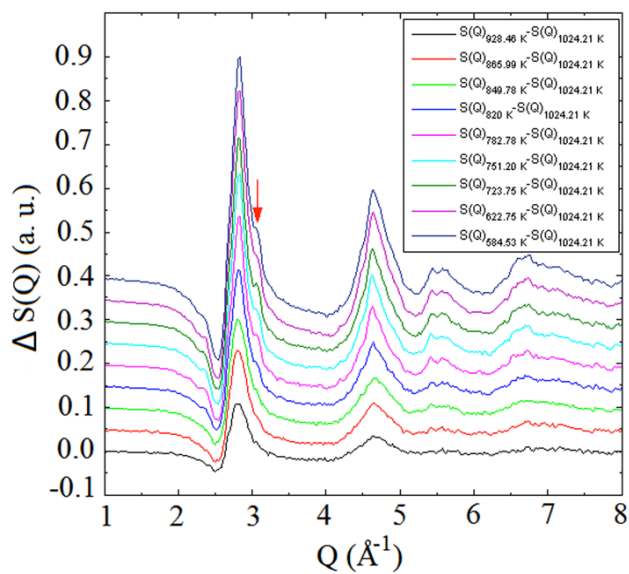

**Fig. S1.** The $\Delta S(Q)$ curves at different temperatures, calculated by subtracting the S(Q) at 1024.21 K. The red arrow indicates the position of a Bragg peak.

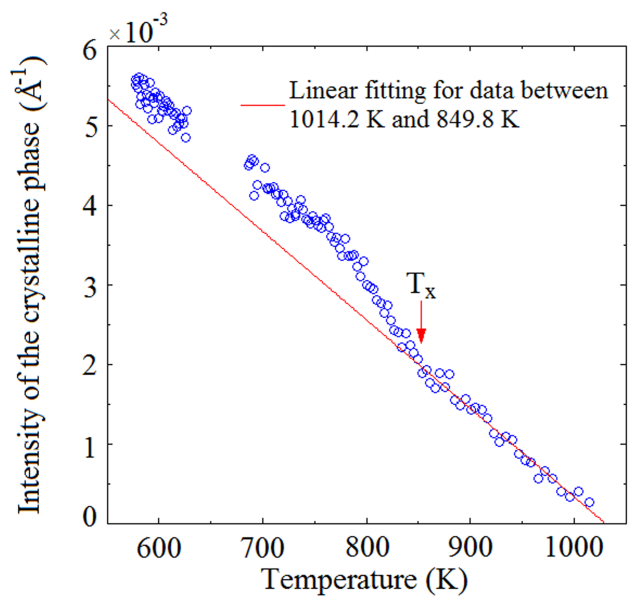

**Fig. S2.** The integrated intensity of the identified crystalline peak in ΔS(Q), over a Q-range of 3.0347-3.0601 Å$^{-1}$. The $T_x$ indicated by a red arrow shows the onset temperature of crystallization. The red line is a linear fit of data above $T_x$.

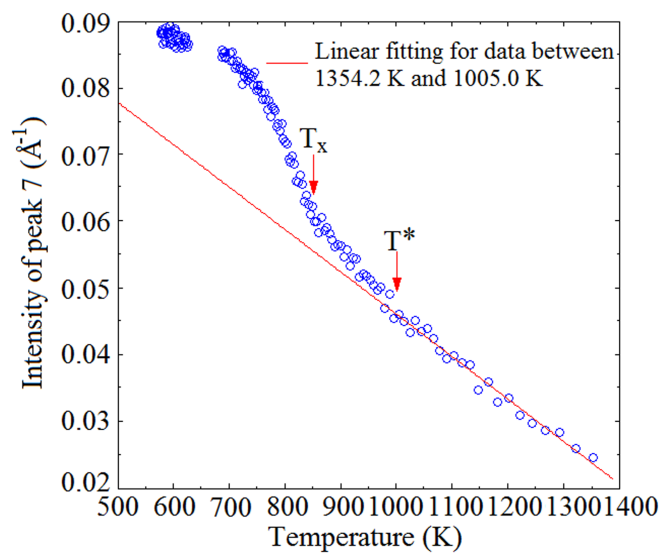

**Fig. S3.** The integrated intensity of the seventh peak in G(r) over the r-range of 15.97-17.01 Å as a function of temperature. Two transition temperature T* and $T_x$, can be readily identified, indicating the growth of extended-range order at large r.